\documentclass[conference]{IEEEtran}
\IEEEoverridecommandlockouts

\usepackage{cite}
\usepackage{amsmath,amssymb,amsfonts}
\usepackage{algorithmic}
\usepackage{graphicx}
\usepackage{textcomp}
\usepackage{xcolor}
\usepackage{booktabs}

\usepackage{cleveref}
\crefname{figure}{Fig.}{Figs.}
\Crefname{figure}{Fig.}{Figs.}
\crefname{table}{Table}{Tables}
\Crefname{table}{Table}{Tables}
\crefname{section}{Section}{Sections}
\Crefname{section}{Section}{Sections}
\crefname{equation}{}{}
\Crefname{equation}{Equation}{Equations}

\newif\ifshowtodo
\showtodotrue

\usepackage{acro}
\DeclareAcronym{wur}{short = WUR, long = wake-up radio}
\DeclareAcronym{wus}{short = WUS, long = wake-up signal}
\DeclareAcronym{nwus}{short = nWUS, long = neuromorphic wake-up signal}
\DeclareAcronym{ofdm}{short = OFDM, long = orthogonal frequency-division multiplexing}
\DeclareAcronym{ook}{short = OOK, long = on-off keying}
\DeclareAcronym{fft}{short = FFT, long = fast Fourier transform}
\DeclareAcronym{ifft}{short = IFFT, long = inverse fast Fourier transform}
\DeclareAcronym{iruwb}{short = IR-UWB, long = impulse-radio ultra-wideband}
\DeclareAcronym{dc}{short = DC, long = direct current}
\DeclareAcronym{papr}{short = PAPR, long = peak-to-average power ratio}
\DeclareAcronym{snr}{short = SNR, long = signal-to-noise ratio}
\DeclareAcronym{pp}{short = pp, long = percentage points}

\begin{document}
\bstctlcite{bstctl}

\title{OFDM-Based Neuromorphic Wake-Up Signal}

\author{\IEEEauthorblockN{Gaspard Goupy}
\IEEEauthorblockA{\textit{Nokia Bell Labs} \\
Espoo, Finland \\
gaspard.goupy@nokia-bell-labs.com}
\and
\IEEEauthorblockN{Elias Raninen}
\IEEEauthorblockA{\textit{Nokia Bell Labs} \\
Espoo, Finland \\
elias.raninen@nokia-bell-labs.com}
\and
\IEEEauthorblockN{Dani Korpi}
\IEEEauthorblockA{\textit{Nokia Bell Labs} \\
Espoo, Finland \\
dani.korpi@nokia-bell-labs.com}
}

\maketitle

\begin{abstract}
Wake-up radios (WURs) reduce energy consumption by continuously monitoring the channel for a wake-up signal (WUS) while the main receiver remains asleep.
Neuromorphic computing offers a promising approach to ultra-low-power WUS detection through sparse, event-driven processing.
However, realizing these benefits requires spike-like inputs, which conventional orthogonal frequency-division multiplexing (OFDM) waveforms do not naturally provide.
To enable neuromorphic WUS detection in OFDM-based systems, we introduce a neuromorphic wake-up signal (nWUS) that can be generated by a standard OFDM transmitter.
The proposed design superposes frequency-domain vectors with linear phase progressions across the OFDM subcarriers, producing a time-domain waveform with sparse amplitude peaks at predefined positions.
These peaks encode the wake-up signature in a spike-like representation that facilitates neuromorphic processing at the receiver.
Building on this structure, we then propose a noncoherent neuromorphic detector that encodes received peaks into spikes and performs sparse correlation only upon spike events.
Simulations over a 3GPP urban macro channel show that the proposed approach outperforms on-off keying (OOK) baselines in detection probability while achieving up to $43.5\times$ lower computational cost.
During idle periods, event-driven processing further reduces correlation operations by $86\%$ relative to nWUS reception.
\end{abstract}

\begin{IEEEkeywords}
wake-up radio, neuromorphic computing, low-power wake-up signal, noncoherent detection, OFDM
\end{IEEEkeywords}

\section{Introduction}

A \ac{wur}~\cite{piyare2017,mercier2022} is an auxiliary low-power receiver that continuously monitors the channel for a specific \ac{wus} while the main receiver is asleep.
\Acp{wur} can reduce energy consumption by waking up the main receiver only when needed, making them particularly attractive for energy-constrained devices~\cite{piyare2017,mercier2022,deng2019}.
To provide meaningful energy savings, both the \ac{wur} and the \ac{wus} should be designed to minimize the processing required for continuous monitoring and detection~\cite{piyare2017,mercier2022,tr38869}.
At the same time, \ac{ofdm} is expected to remain a key waveform technology for 6G.
Compatibility with standard \ac{ofdm} transmitters is thus desirable for \ac{wus} design~\cite{mazloum2020,tr38869,hoglund2024,wagner2025}.

Neuromorphic computing~\cite{schuman2017,davies2021} provides a promising approach for ultra-low-power \acp{wur}~\cite{jouni2022,marthe2023}.
Inspired by biological neural networks, neuromorphic systems represent information as discrete binary events called spikes.
They process these events in a sparse and event-driven manner~\cite{schuman2017}, i.e., with computation occurring upon spike events.
When implemented on neuromorphic hardware~\cite{davies2021}, this can substantially reduce the computational cost and associated energy consumption compared with conventional dense processing, especially during idle periods (when the \ac{wur} receives only noise).
Recent studies have demonstrated the potential of neuromorphic \acp{wur} for energy-efficient \ac{wus} detection~\cite{jouni2022,jouni2023,marthe2023}; for example, a 1.2~nW analog implementation was shown in~\cite{jouni2022}.
However, efficient neuromorphic processing requires inputs represented as discrete spike events~\cite{davies2021}.
Conventional \ac{ofdm} waveforms lack this spike-like structure, which limits their direct use with neuromorphic \acp{wur}.

\begin{figure*}[t]
  \centering
  \includegraphics[width=\textwidth]{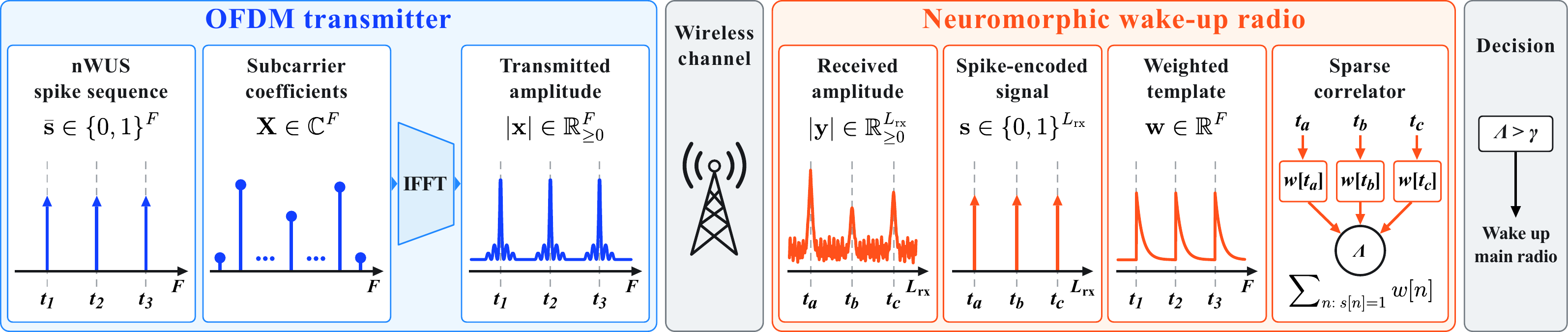}
  \caption{Overview of the proposed nWUS scheme, comprising OFDM-based nWUS generation and neuromorphic detection. The OFDM transmitter maps the nWUS signature, defined as a spike sequence, to subcarrier coefficients. OFDM modulation then produces sparse amplitude peaks at the spike positions. The receiver encodes the received amplitude peaks into spike events and performs sparse correlation with a template derived from the nWUS signature.}
  \label{fig:nwus-pipeline}
\end{figure*}

In this paper, we introduce a \ac{nwus} that can be generated by a standard \ac{ofdm} transmitter, as well as a neuromorphic detector that exploits the spike-like structure of the \ac{nwus} for efficient \ac{wus} detection.
The overall method is illustrated in \cref{fig:nwus-pipeline}.
The main contributions of this paper are as follows.
\begin{itemize}
\item We propose an \ac{ofdm}-based \ac{nwus} whose time-domain waveform uses spike-like amplitude peaks at configurable positions to encode the wake-up signature, enabling efficient neuromorphic processing at the receiver.
\item We propose a noncoherent neuromorphic detector that encodes the received amplitude peaks into spikes and performs sparse correlation using a weighted template derived from the target \ac{nwus}. The correlator processes only spike events, reducing computation relative to dense correlation and further lowering activity during idle periods when fewer spikes are generated.
\item We evaluate the \ac{nwus} over a 3GPP urban macro channel and compare it with two \ac{ook} baselines of comparable and tenfold duration. Under equal average transmit power, the \ac{nwus} improves mean detection probability by $18.6$ and $3.3$~\ac{pp} while reducing counted operations by $2.5\times$ and $43.5\times$, respectively. During idle periods, correlation operations are further reduced by $86\%$ relative to \ac{nwus} reception.
\end{itemize}
Overall, this work contributes toward enabling ultra-low-power neuromorphic \acp{wur} for \ac{ofdm}-based wireless systems.

\section{Related Work}

\subsection{Low-Power Wake-Up Signals}
Low-power \acp{wus} are designed to enable reliable detection with minimal processing at the \ac{wur}~\cite{mercier2022,tr38869,wagner2025}.
To this end, the \ac{wus} should support noncoherent time-domain detection, avoiding carrier synchronization, \ac{fft} operations, and frequency-domain calibration~\cite{deng2020,tr38869,wagner2025}.
Existing approaches mainly use \ac{ook}, which can be implemented using \ac{ofdm} transmitters and detected with simple noncoherent envelope-based receivers~\cite{deng2019,deng2020,mazloum2020,tr38869,hoglund2024,wagner2025,pitaval2025}.
However, in \ac{ook}, bits are typically transmitted sequentially, each represented by an ON/OFF symbol occupying a dedicated time interval.
This can lead to long \acp{wus} and continuous, dense receiver processing, thereby limiting the energy efficiency achievable at the \ac{wur}~\cite{tr38869,hoglund2024}.
Prior work has used energy concentration in \ac{ook} to improve detection robustness~\cite{pitaval2025}, but combining it with sparsity to lower receiver processing costs has not been investigated.

\subsection{Neuromorphic Wake-Up Radios}
Neuromorphic computing has recently been explored for \ac{wur} applications~\cite{jouni2022,jouni2023,marthe2023,marthe2025,chen2024tsp,chen2024spawc}.
Existing work mainly considers \ac{ook} as the \ac{wus} waveform~\cite{jouni2022,jouni2023,marthe2023}.
\Ac{ook} can be represented as spike events based on the presence or absence of received energy, but its waveform is neither inherently sparse nor spike-like.
In parallel, \acp{wur} have been integrated into neuromorphic communication architectures based on \ac{iruwb}~\cite{chen2024tsp,chen2024spawc}.
\Ac{iruwb} signals provide sparse temporal impulses well suited to neuromorphic processing, but are not compatible with conventional \ac{ofdm} transmitters.
To our knowledge, \ac{ofdm}-based \ac{wus} designs that exploit sparse, spike-like time-domain waveforms for neuromorphic detection remain unexplored.

\section{Background}

\textbf{Notation:} Bold symbols denote column vectors, and $a[n]$ denotes the $n$-th entry of $\mathbf{a}$. Vector entries and FFT bins are indexed starting at $1$, whereas detection lags start at $0$. The imaginary unit is denoted by $\mathrm{j}$, with $\mathrm{j}^2=-1$.

\subsection{System Model}
\label{sec:system_model}

We consider an \ac{ofdm}-based downlink wake-up scenario.
At the transmitter, a \ac{wus} is mapped onto an \ac{ofdm} symbol and transmitted over the air.
At the receiver, the \ac{wur} searches for a known \ac{wus} directly in the time-domain baseband signal.

Let $\mathbf{x}\in\mathbb{C}^{L_{\mathrm{tx}}}$ denote the transmitted baseband \ac{wus} waveform comprising $L_{\mathrm{tx}}$ time-domain samples.
For an \ac{ofdm}-based \ac{wus}, a conventional \ac{ofdm} transmitter with \ac{fft} size $F$ converts the frequency-domain subcarrier coefficients $\mathbf{X}\in\mathbb{C}^{F}$ into time-domain samples via the $F$-point \ac{ifft}
\begin{equation}
x[n] \triangleq \frac{1}{F} \sum_{k=1}^{F} X[k]
\exp\left(\mathrm{j}\frac{2\pi (k-1)(n-1)}{F}\right),
\label{eq:ifft}
\end{equation}
where $n=1,\ldots,F$, giving $L_{\mathrm{tx}}=F$; the cyclic prefix is omitted for simplicity.

Let $\mathbf{y}\in\mathbb{C}^{L_{\mathrm{rx}}}$ denote the received time-domain baseband signal, where $L_{\mathrm{rx}}$ is the observation length.
For a time-varying finite-impulse-response channel, the received samples are given by
\begin{equation}
y[n] = \sum_{m=1}^{L_h} h[n,m]\,x[n-m+1] + z[n],
\label{eq:rx}
\end{equation}
where $L_h$ is the channel impulse response length, $z[n]$ is additive noise, and $h[n,m]$ is the $m$-th tap of the discrete-time baseband channel impulse response at sample $n$; $x[n]=0$ outside $1\leq n\leq L_{\mathrm{tx}}$.
The \ac{wur} performs noncoherent detection on the received amplitude $\mathbf{r}\triangleq|\mathbf{y}|\in\mathbb{R}_{\ge0}^{L_{\mathrm{rx}}}$.

Because the arrival time is unknown, the observation at the \ac{wur} is longer than the transmitted waveform: one detection attempt processes $L_{\mathrm{rx}} = L_{\mathrm{tx}}+N_{\mathrm{lag}}-1$ samples and evaluates $N_{\mathrm{lag}}$ lags spaced by one sample.
Lag $\ell=0,\ldots,N_{\mathrm{lag}}-1$ corresponds to the window of samples $\ell+1,\ldots,\ell+L_{\mathrm{tx}}$.

\subsection{Problem Formulation}
\label{sec:problem_formulation}

We primarily consider a \ac{wus} identification problem in which the receiver must distinguish a target \ac{wus} from other valid \acp{wus}.
We focus on this setting because it can be more challenging than detection against noise only (both hypotheses contain a valid transmitted signal).
Detection is formulated as a binary hypothesis test per attempt: hypothesis $\mathcal{H}_1$ denotes reception of the target \ac{wus} and $\mathcal{H}_0$ denotes reception of a \ac{wus} from the same family with a different signature (referred to as a non-target \ac{wus}).
The \ac{wur} computes a score $v_\ell$ for each lag window and compares the maximum $\Lambda\triangleq\max_{0\leq\ell\leq N_{\mathrm{lag}}-1} v_\ell$ with a decision threshold $\gamma\in\mathbb{R}$, selecting $\mathcal{H}_1$ if $\Lambda>\gamma$.
The detection probability $P_{\mathrm{d}}\triangleq\Pr\{\Lambda>\gamma\mid\mathcal{H}_1\}$ and the false-identification probability $P_{\mathrm{fi}}\triangleq\Pr\{\Lambda>\gamma\mid\mathcal{H}_0\}$ are therefore defined per attempt, i.e., over the whole lag search rather than per window, and $\gamma$ is calibrated accordingly.

In addition to this identification task, we also consider detection against noise only by replacing $\mathcal{H}_0$ with $\mathcal{H}_{\mathrm{noise}}$, under which the received signal in \cref{eq:rx} reduces to $y[n]=z[n]$.
The corresponding per-attempt false-alarm probability is $P_{\mathrm{fa}}\triangleq\Pr\{\Lambda>\gamma\mid\mathcal{H}_{\mathrm{noise}}\}$.

\subsection{On-Off Keying Baseline}
\label{sec:ook}

We consider an \ac{ook}-based \ac{wus}, denoted OOK-WUS, whose signature is a binary sequence $\bar{\mathbf{b}}\in\{0,1\}^{N}$.
A bit ``1'' is transmitted as an active (ON) interval, whereas a bit ``0'' is transmitted as an inactive (OFF) interval.
Each bit spans $L_{\mathrm{b}}\in\mathbb{Z}_{>0}$ time samples, resulting in a waveform of length $L_{\mathrm{tx}} = N L_{\mathrm{b}}$ samples.
We generate the waveform directly in the time domain as $x[n]=\bar{b}[\lceil n/L_{\mathrm{b}}\rceil]$, without transmit filtering, to obtain an idealized reference.
\ac{ofdm}-based realizations can also be used~\cite{deng2020} but introduce additional waveform design choices beyond the scope of this baseline.

For noncoherent detection with unknown bit boundaries, the received squared amplitude is integrated over sliding intervals of $L_{\mathrm{b}}$ samples, yielding the energy sequence $u[p]\triangleq\sum_{n=p}^{p+L_{\mathrm{b}}-1} r[n]^2$.
For each lag $\ell$, the corresponding bit-energy vector $\mathbf{e}_\ell\in\mathbb{R}_{\ge0}^{N}$ has entries $e_\ell[i]\triangleq u[\ell+(i-1)L_{\mathrm{b}}+1]$.
To improve identification performance, $\mathbf{e}_\ell$ is mean-centered and normalized to unit norm, giving $\hat{\mathbf{e}}_\ell$ (if the centered vector has zero norm, $\hat{\mathbf{e}}_\ell = \mathbf{0}$).
The receiver then correlates it with the target \ac{ook} sequence $\bar{\mathbf{b}}$, producing the per-lag score $v_\ell\triangleq\hat{\mathbf{e}}_\ell^\top\bar{\mathbf{b}}$.
Last, the decision follows \cref{sec:problem_formulation}.

\section{Neuromorphic Wake-Up Signal}
\label{sec:nwus}

We seek a \ac{wus} that is compatible with conventional \ac{ofdm} transmitters and has a sparse, spike-like time-domain structure, enabling efficient neuromorphic processing at the receiver; we introduce the \ac{nwus} to satisfy these requirements.

\subsection{Signature Definition}

The \ac{nwus} signature is a length-$F$ spike sequence in which $N_{\mathrm{s}}$ spike events occur at predefined time indices $\mathcal{T}\triangleq\{t_i\}_{i=1}^{N_{\mathrm{s}}}$, ordered such that $1\leq t_1<\cdots<t_{N_{\mathrm{s}}}\leq F$.
Equivalently, the \ac{nwus} signature can be defined as a binary sequence $\bar{\mathbf{s}}\in\{0,1\}^{F}$ with entries $\bar{s}[n]=1$ if $n\in\mathcal{T}$ and $\bar{s}[n]=0$ otherwise.
We impose a minimum spacing $\Delta_{\min}\in\mathbb{Z}_{>0}$ between consecutive spikes, such that $t_{i+1}-t_i\geq\Delta_{\min}$.
The signature is fixed and known to the \ac{wur} during operation; different \acp{wur} can be assigned different signatures.

Two main design considerations guide the choice of $\mathcal{T}$.
First, $N_{\mathrm{s}}$ controls the sparsity of the signature and provides a trade-off between detection robustness and receiver-side computational cost, since more spikes require more operations.
Second, $\Delta_{\min}$ should be large enough to keep the amplitude peaks corresponding to adjacent spikes distinguishable in the generated time-domain waveform.

\subsection{Waveform Construction}

We consider an \ac{nwus} occupying one \ac{ofdm} symbol with \ac{fft} size $F$ and $D \leq F$ active subcarriers.
Let $\mathcal{B}\triangleq\{\kappa_d\}_{d=1}^{D}\subseteq\{1,\ldots,F\}$ denote the indices of the active subcarriers, which form a contiguous frequency band.
Our goal is to select the subcarrier coefficients so that, after the \ac{ifft}, the time-domain waveform exhibits sparse amplitude peaks at the predefined indices in $\mathcal{T}$.

For each spike position $t_i \in \mathcal{T}$, we construct a subcarrier coefficient vector $\mathbf{X}^{(i)}_{\mathcal{B}}\in\mathbb{C}^{D}$ with entries
\begin{equation}
X^{(i)}_{\mathcal{B}}[d] \triangleq \exp\left(-\mathrm{j}\frac{2\pi(\kappa_d-1)(t_i-1)}{F}\right).
\label{eq:nwus_single_spike}
\end{equation}
The linear phase progression makes the active-subcarrier contributions add coherently at sample $t_i$ after the \ac{ifft}, producing a spike-like amplitude peak.
To obtain all spikes in one \ac{ofdm} symbol, we sum these vectors to form the \ac{nwus} subcarrier coefficient vector $\mathbf{X}_{\mathcal{B}}\triangleq\sum_{i=1}^{N_{\mathrm{s}}}\mathbf{X}^{(i)}_{\mathcal{B}}\in\mathbb{C}^{D}$.
Equivalently, $\mathbf{X}_{\mathcal{B}}$ can be formed by applying an $F$-point \ac{fft} to the spike sequence $\bar{\mathbf{s}}$ and retaining the coefficients at bins $\kappa_d\in\mathcal{B}$.
We emphasize that our contribution lies not in generating time-domain peaks through coherent phase alignment, but in using these peaks and their positions as a sparse \ac{wus} signature for efficient neuromorphic processing.

We construct the full frequency-domain \ac{ofdm} symbol $\mathbf{X}\in\mathbb{C}^{F}$ by setting $X[\kappa_d]=X_{\mathcal{B}}[d]$ for $d=1,\ldots,D$ and all remaining bins to zero.
Here, we choose $\mathcal{B}$ to occupy a contiguous band centered around the \ac{dc} subcarrier.
An $F$-point \ac{ifft} is then applied to $\mathbf{X}$ via \cref{eq:ifft}, producing a time-domain waveform with sparse amplitude peaks at the predefined indices in $\mathcal{T}$.
When all $F$ subcarriers are active, the \ac{ifft} exactly reconstructs the \ac{nwus} spike sequence $\bar{\mathbf{s}}$, with each spike occupying a single time sample.
For $D<F$, the band-limited spectrum produces a Dirichlet-kernel-shaped pulse around each spike position, with a broader main lobe and deterministic sidelobes.
Provided that $\Delta_{\min}$ is sufficiently large, the peaks should remain distinguishable and their locations should be preserved.
The \ac{dc} subcarrier is kept active since the \ac{wur} operates in the time domain, although \ac{dc} nulling is also supported at the cost of a small deterministic distortion of the time-domain waveform.
The entire \ac{ofdm} symbol is reserved for the \ac{nwus} (i.e., not shared with data) to preserve the intended temporal structure.

Last, the waveform proceeds through the rest of the \ac{ofdm} transmission chain.
A cyclic prefix is not needed by the \ac{nwus} scheme itself, but it can be included when required by the \ac{ofdm} configuration.
In that case, the spike positions should satisfy $t_{N_{\mathrm{s}}}\leq F-L_{\mathrm{cp}}$, where $L_{\mathrm{cp}}$ denotes the cyclic-prefix length, to exclude the peak centers from the samples copied into the prefix.
Overall, the \ac{nwus} modifies only the frequency-domain coefficients given to the \ac{ofdm} modulator and thus requires no changes to the \ac{ofdm} transmitter.

\subsection{Information Capacity}
\label{sec:nwus_capacity}
The \ac{nwus} can encode information through the number of spikes and their positions.
For an \ac{fft} size $F$, a fixed number of spikes $N_{\mathrm{s}}$, and the minimum-spacing constraint $t_{i+1}-t_i\geq\Delta_{\min}$, the number of distinct spike sequences is
\begin{equation}
N_{\mathrm{seq}}\triangleq \binom{F-(\Delta_{\min}-1)(N_{\mathrm{s}}-1)}{N_{\mathrm{s}}},
\label{eq:nwus_num_sequences}
\end{equation}
where feasibility requires $F\geq1+(N_{\mathrm{s}}-1)\Delta_{\min}$.
Hence, one \ac{nwus} symbol can carry at most $C_{\mathrm{nWUS}}\triangleq\left\lfloor\log_2 N_{\mathrm{seq}}\right\rfloor$ bits.
In practice, the usable capacity may be lower due to, for example, ambiguity between shifted signatures under the lag search or limited robustness of some sequences to channel impairments.

\section{Neuromorphic Detector}
\label{sec:detector}

We propose a noncoherent neuromorphic detector to efficiently process the \ac{nwus} at the receiver.

\subsection{Spike Encoding}

The detector first encodes received amplitude peaks into spike events, converting the continuous signal into a sparse representation that reduces subsequent computational cost.
In this work, we use a simple fixed-threshold level-crossing encoder.
Encoding is performed once per detection attempt by comparing each of the $L_{\mathrm{rx}}$ received amplitude samples with an encoder threshold $\eta \in \mathbb{R}_{>0}$.
The spike-encoded signal $\mathbf{s}\in\{0,1\}^{L_{\mathrm{rx}}}$ is defined as
\begin{equation}
s[n]\triangleq
\begin{cases}
1, & r[n]\geq\eta\ \text{and}\ r[n-1]<\eta,\\
0, & \text{otherwise},
\end{cases}
\label{eq:level_crossing}
\end{equation}
so one spike event is emitted on each rising crossing rather than on every sample above the threshold.
The threshold balances detection robustness and computational cost: it must be low enough to detect \ac{nwus} peaks but high enough to avoid generating too many spikes during idle periods (when only noise is present).
To control this trade-off, we define an average spike budget $K\in \mathbb{R}_{>0}$ over the $L_{\mathrm{rx}}$-sample observation window and select $\eta$ so that the empirical average spike count under $\mathcal{H}_1$ matches this budget.
Setting $K>N_{\mathrm{s}}$ allows more spikes than transmitted \ac{nwus} peaks, which in practice can improve the robustness of the subsequent detector.

Limitations of this spike encoder include its reliance on absolute amplitude levels and the need to calibrate $\eta$ prior to operation.
However, the proposed \ac{nwus} and correlator are agnostic to the encoder.
Adaptive encoders, such as constant-false-alarm-rate schemes, could reduce calibration requirements and improve robustness to changing conditions.
Also, spike encoding can be performed in the analog domain~\cite{jouni2022}, avoiding high-rate sampling and analog-to-digital conversion.

\subsection{Sparse Correlator}
\label{sec:correlator}

A sparse correlator matched to the target \ac{nwus} signature then processes the spike-encoded signal.
The correlator follows two key neuromorphic principles~\cite{schuman2017}: discrete spike inputs and computation triggered only by spikes.
The correlator compares windows of $F$ consecutive spike-encoded samples (matching the transmitted \ac{nwus} length) with a template represented by a real-valued weight vector $\mathbf{w}\in\mathbb{R}^{F}$.
Since the \ac{nwus} may arrive at an unknown position, the correlation is evaluated at every lag $\ell=0,\ldots,N_{\mathrm{lag}}-1$:
\begin{equation}
v_\ell\triangleq\sum_{n=1}^{F} s[\ell+n]\,w[n]
=\sum_{\substack{1\leq n\leq F\\s[\ell+n]=1}} w[n],
\label{eq:score}
\end{equation}
and the maximum score $\Lambda = \max_{0\leq\ell\leq N_{\mathrm{lag}}-1}v_\ell$ is kept.
Because only active spike positions contribute to \cref{eq:score}, the computational cost of the correlator scales with the number of spikes rather than with the window length.
This reduction in computation translates into lower energy usage on hardware that exploits sparsity, such as neuromorphic processors~\cite{schuman2017,davies2021}.

The weighted template $\mathbf{w}$ is constructed directly from the target \ac{nwus} signature.
All weights are first initialized to a minimum value $w_{\min}$.
Then, for each spike at position $t_i$, the weights are defined for $t_i \leq n < t_{i+1}$ as
\begin{equation}
w[n] \triangleq w_{\min} + \left(w_{\max}-w_{\min}\right) \exp\left(-\frac{n-t_i}{\tau}\right),
\label{eq:init}
\end{equation}
where $w_{\max}$ is the maximum value, $\tau>0$ is the decay constant in samples, and $t_{N_{\mathrm{s}}+1}\triangleq F+1$.
Thus, each spike sets the weight to $w_{\max}$, which then decays toward $w_{\min}$.

The decay provides tolerance to local temporal distortions within an aligned window, such as band-limiting and multipath smearing.
A spike received slightly after its expected position still contributes positively to the score.
The decay constant $\tau$ controls the trade-off between robustness and temporal selectivity.
This tolerance is distinct from the lag search, which absorbs the global delay.
Moreover, detection does not require exact recovery of the transmitted spike sequence; the correlation score can remain discriminative when some spikes are missed, temporally shifted, or accompanied by additional unintended spikes.
Last, the sparse correlator is hardware-agnostic, supporting implementations in conventional digital circuits, analog circuits, and neuromorphic hardware.

\section{Simulation Results}
\label{sec:results}

We evaluate the proposed \ac{nwus} scheme and compare its performance with two OOK-WUS baselines.

\subsection{Setup}
\label{sec:sim_setup}

\paragraph{nWUS}
The \ac{nwus} uses one \ac{ofdm} symbol with an \ac{fft} size of $F=1024$, $D=870$ active subcarriers ($85$\% occupancy, leaving the remaining subcarriers for guard bands), a subcarrier spacing of $15$~kHz, and no cyclic prefix (not required here).
The resulting sampling rate is $15.36$~MHz, and the \ac{nwus} duration is $66.67$~\textmu s.
We set the number of spikes to $N_{\mathrm{s}}=5$ and the minimum spike spacing to $\Delta_{\min}=20$ samples; the \ac{nwus} signature is generated randomly subject to these sparsity and spacing constraints.
The correlator parameters are $\tau=3$, $w_{\min}=-0.1$, $w_{\max}=1$, and the encoder spike budget is set to $K=14$ spikes per observation.
The parameters $N_{\mathrm{s}}$, $\Delta_{\min}$, $K$, $\tau$, and $w_{\min}$ were selected via random search on independent data to balance performance and sparsity.

\paragraph{OOK-WUS}
We use two OOK-WUS baselines from \cref{sec:ook} to provide complementary comparisons with the \ac{nwus}.
The first contains $N=1024$ bits, matching the length of the \ac{nwus} signature.
Each bit spans $L_{\mathrm{b}}=10$ samples, a duration selected to approach the detection reliability of the \ac{nwus}, yielding ten times its airtime ($666.67$~\textmu s).
The second contains $N=42$ bits, such that this waveform and the \ac{nwus} support approximately the same number of distinct \ac{wus} signatures (see \cref{sec:nwus_capacity}).
Here, $L_{\mathrm{b}}=25$ is selected to closely match the duration of the \ac{nwus} (OOK-WUS duration of $68.36$~\textmu s), providing a comparison with similar information capacity and airtime.
Both OOK-WUS baselines use the same $15.36$~MHz sampling rate as the \ac{nwus}.
The OOK-WUS signatures are randomly generated with independent equiprobable bits.

\paragraph{Channel model}
We use Sionna 2.0~\cite{sionna} to simulate the 3GPP TR~38.901 urban macro channel~\cite{tr38901}.
All waveforms are evaluated over the same single-input single-output downlink at $3.5$~GHz.
For each realization, user locations, heights ($[1.5,22.5]$~m), indoor/outdoor conditions (indoor probability of $0.5$), and speeds ($[0,30]$~m/s) are randomly generated.
The channel response is normalized to unit energy, and independent complex AWGN is added.
The receivers search $N_{\mathrm{lag}}=60$ sample-spaced lags to account for channel-induced delays; this gives $L_{\mathrm{rx}}=1{,}083$ for the \ac{nwus}, $1{,}109$ for the $42$-bit OOK-WUS, and $10{,}299$ for the $1024$-bit OOK-WUS.
No random timing offset is introduced beyond the delay induced by the channel.
All waveforms are scaled to the same average transmit power before channel propagation to achieve the target \ac{snr}.\footnote{We define \ac{snr} as the ratio of the average transmitted signal power per sample (before the channel) to the noise power per complex sample.}

\paragraph{Evaluation protocol}
We evaluate performance through Monte Carlo simulations of the detection problem defined in \cref{sec:problem_formulation}.
Target and non-target \acp{wus} use identical parameters and differ only in their signatures, defined by the spike positions for the \ac{nwus} and the bit sequence for the OOK-WUS.
The target signature is fixed, whereas a non-target signature is drawn independently for every $\mathcal{H}_0$ attempt.
For the \ac{nwus}, we first calibrate a single encoder threshold $\eta$ by matching the empirical mean spike count to the budget $K$ over $10^4$ independent $\mathcal{H}_1$ attempts with random \acp{snr} in the evaluated range.
After this preliminary step, the \ac{nwus} and the OOK-WUS baselines are evaluated using the same procedure.
At each \ac{snr} point, we calibrate the decision threshold $\gamma$ from $10^5$ independent $\mathcal{H}_0$ attempts by selecting the empirical quantile corresponding to the target false-identification probability $P_{\mathrm{fi}}=10^{-2}$.
Using the resulting threshold, we estimate the detection probability $P_{\mathrm{d}}$ from $10^5$ independent $\mathcal{H}_1$ attempts and verify the achieved $P_{\mathrm{fi}}$ on a separate set of $10^5$ independent $\mathcal{H}_0$ attempts.
For the secondary noise-only problem, we replace $\mathcal{H}_0$ with $\mathcal{H}_{\mathrm{noise}}$ throughout this procedure: $\gamma$ is calibrated using noise-only attempts for a target $P_{\mathrm{fa}}=10^{-2}$, while all other settings remain unchanged.
To assess sensitivity to the \ac{wus} signature, we repeat the experiment for five random seeds, each defining a different target signature.
The reported results show the mean across these five runs.

\subsection{Detection Performance}
\label{sec:performance}

\begin{figure}[t]
  \centering
  \includegraphics[width=0.999\columnwidth]{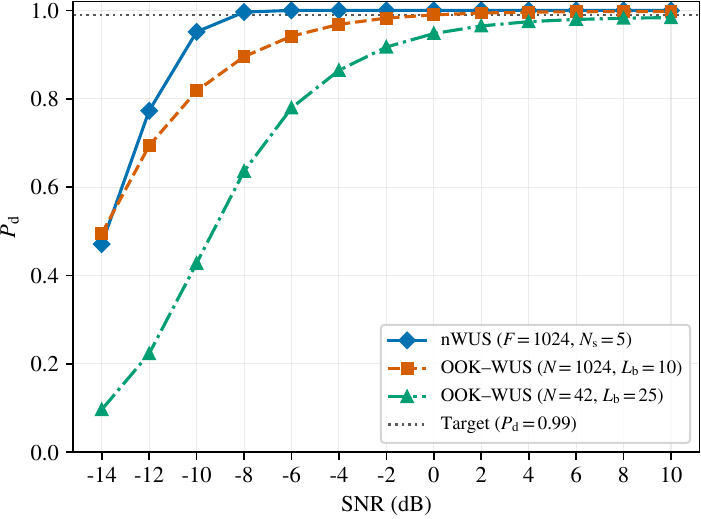}
  \caption{Detection probability $P_{\mathrm{d}}$ versus \ac{snr} under equal average transmit power, with the decision threshold calibrated to achieve $P_{\mathrm{fi}}=10^{-2}$.}
  \label{fig:pd_power}
\end{figure}

\Cref{fig:pd_power} shows the detection probability as a function of \ac{snr} for the proposed \ac{nwus} and the two OOK-WUS baselines, all evaluated under equal average transmit power.
We use $P_{\mathrm{d}}=0.99$ as the target detection reliability.
The achieved $P_{\mathrm{fi}}$ remains in the range $[0.93,\,1.11]\times10^{-2}$.

Compared with the $42$-bit OOK-WUS, which has a similar waveform duration and information capacity, the \ac{nwus} achieves a substantially higher $P_{\mathrm{d}}$ over the entire \ac{snr} range, with a gain of $18.6$~\ac{pp} averaged over all evaluated \ac{snr} points.
The $1024$-bit OOK-WUS is ten times longer than the \ac{nwus} and therefore transmits ten times more energy at the same average transmit power.
Despite this advantage, it is outperformed by the \ac{nwus}, which achieves a mean $P_{\mathrm{d}}$ gain of $3.3$~\ac{pp}.
The \ac{nwus} reaches the target $P_{\mathrm{d}}$ at an \ac{snr} of approximately $-8$~dB, compared with approximately $0$~dB for the $1024$-bit OOK-WUS, while the $42$-bit OOK-WUS remains below the target throughout the evaluated \ac{snr} range.
Overall, these results show that the proposed \ac{nwus} achieves higher detection performance than OOK-WUS with comparable or lower airtime.
We attribute the performance gain primarily to two factors: (i) concentrating the available \ac{wus} energy into sparse temporal peaks, which improves spike recovery at the receiver, and (ii) using a weighted correlator template tailored to spike sequences, which also improves robustness to local, channel-induced spike shifts.

Finally, we evaluate detection against noise only.
In this setting, the \ac{nwus} reaches $P_{\mathrm{d}}=0.99$ at $-10$~dB, compared with $-6$~dB for the $1024$-bit OOK-WUS and $0$~dB for the $42$-bit OOK-WUS.
These results show that the \ac{nwus} retains its detection advantage over both baselines when distinguishing the target \ac{wus} from noise only.

\subsection{Computational Cost}
\label{sec:cost}

A main motivation for the \ac{nwus} is to reduce the computational cost at the receiver through sparse, event-driven computing.
We assess this by approximating the number of operations required for one detection attempt.
For the \ac{nwus} detector, the level-crossing encoder performs one comparison per sample, giving $L_{\mathrm{rx}}$ preprocessing operations.
The sparse correlator performs one addition per spike in each lag window.
Its total cost per detection attempt is therefore $N_{\mathrm{lag}}$ times the average number of spikes per window, measured as $13.4$ under $\mathcal{H}_1$ and $1.9$ under $\mathcal{H}_{\mathrm{noise}}$.
For the OOK-WUS detector, the running sum providing the bit energies requires $2L_{\mathrm{rx}}$ preprocessing operations (one addition and one subtraction per sample).
The dense correlator requires $N \times N_{\mathrm{lag}}$ additions; it is kept dense (without zero skipping) to isolate the benefits of sparse computation on appropriate hardware.
We omit the mean-centering and unit-norm normalization used to improve performance because their computational cost is strongly implementation-dependent and they have no direct counterpart in the \ac{nwus} detector.
Each comparison, addition, or subtraction is counted as one operation, as their energy costs are comparable~\cite{manzouri2022}.

\begin{table}[t]
\caption{Estimated computational cost per detection attempt.}
\label{tab:cost}
\centering
\footnotesize
\setlength{\tabcolsep}{4pt}
\begin{tabular}{lrrrr}
\toprule
& \multicolumn{3}{c}{Operations} & \\
\cmidrule(lr){2-4}
Scheme & Preproc. & Corr. & Total & Rel. cost \\
\midrule
nWUS ($\mathcal{H}_{\mathrm{noise}}$) & $1{,}083$ & $114$ & $1{,}197$ & $0.6\times$ \\
nWUS ($\mathcal{H}_1$) & $1{,}083$ & $804$ & $1{,}887$ & $1\times$ \\
$42$-bit OOK-WUS ($\mathcal{H}_1$) & $2{,}218$ & $2{,}520$ & $4{,}738$ & $2.5\times$ \\
$1024$-bit OOK-WUS ($\mathcal{H}_1$) & $20{,}598$ & $61{,}440$ & $82{,}038$ & $43.5\times$ \\
\bottomrule
\end{tabular}
\end{table}

\Cref{tab:cost} reports the number of counted operations for each scheme.
At comparable airtime and information capacity, the $42$-bit OOK-WUS requires $2.5\times$ as many counted operations as the \ac{nwus} under $\mathcal{H}_1$.
This ratio increases to $43.5\times$ for the $1024$-bit OOK-WUS, whose tenfold airtime further increases the processing cost.
During noise-only periods ($\mathcal{H}_{\mathrm{noise}}$), the \ac{nwus} correlator performs approximately $86\%$ fewer operations than under $\mathcal{H}_1$, while the OOK-WUS correlator incurs the same cost regardless of received activity.
This illustrates the computational benefits of sparse, event-driven processing.
Together with the findings in \cref{sec:performance}, these results show that the \ac{nwus} combines computational savings with higher detection probability relative to OOK-WUS baselines of both comparable and substantially longer duration.
We note that the reported operation counts should be interpreted as coarse proxies for relative computational cost and associated receiver energy, as they omit additional operations and implementation-dependent costs such as memory access.

\section{Conclusion}

At a high level, both an \ac{nwus} and an OOK-WUS encode a bit-like temporal signature that can be detected using correlation.
Their main difference lies in how this signature is represented in the time-domain waveform.
In an OOK-WUS, active symbols are frequent and extend over multiple time samples, whereas in an \ac{nwus}, active symbols occur sparsely and are localized to one or a few time samples.
This sparse representation enables a temporally compact waveform that concentrates transmit energy into fewer active samples.
At the receiver, energy concentration improves detectability, sparsity reduces the computational cost of correlation, and event-driven processing further reduces this cost during idle periods.
These benefits are obtained while retaining \ac{ofdm} compatibility, making the proposed \ac{nwus} a promising waveform for ultra-low-power neuromorphic \acp{wur} in \ac{ofdm}-based wireless systems.
More broadly, the \ac{nwus} is not tied to the proposed detector and could also benefit existing neuromorphic \ac{wur} architectures that rely on spike-based \ac{wus} signatures.

Realizing these benefits in practical systems requires further investigation of hardware constraints and integration aspects.
In particular, the impact of the high amplitude peaks on power amplifier operation remains unassessed and should be studied in future work.
We will also investigate multiplexing the \ac{nwus} with user data on unused subcarriers to improve spectral efficiency.

\bibliographystyle{IEEEtran}
\bibliography{references}

\end{document}